\documentclass[twocolumn,showpacs,preprintnumbers,amsmath,amssymb,superscriptaddress]{revtex4}
\usepackage{graphicx,dcolumn,bm}

\newcommand{\Sec}[1]{Sec.\,\ref{#1}}

\newcommand{\nl}{\nonumber \\}
\newcommand{\be}{\begin{equation}}
\newcommand{\ee}{\end{equation}}
\newcommand{\bea}{\begin{eqnarray}}
\newcommand{\eea}{\end{eqnarray}}
\newcommand{\bsube}{\begin{subequations}}
\newcommand{\esube}{\end{subequations}}
\newcommand{\Fig}[1]{Fig.\,\ref{#1}}
\newcommand{\Eq}[1]{Eq.\,(\ref{#1})}
\newcommand{\rmS}{{\rm S}}
\newcommand{\rmB}{{\rm B}}
\newcommand{\rmL}{{\rm L}}
\newcommand{\rmR}{{\rm R}}

\newcommand{\rmc}{{\rm c}}
\newcommand{\rmi}{{\rm i}}
\newcommand{\rmd}{{\rm d}}
\newcommand{\alf}{\alpha}
\newcommand{\sgm}{\sigma}
\newcommand{\Omg}{\Omega}
\newcommand{\omg}{\omega}
\newcommand{\Gam}{\Gamma}
\newcommand{\Dlt}{\Delta}
\newcommand{\dlt}{\delta}
\newcommand{\vpl}{\varepsilon}
\newcommand{\epl}{\epsilon}
\newcommand{\upa}{\uparrow}
\newcommand{\dwa}{\downarrow}
\newcommand{\GamL}{\Gamma_{\rm L}}
\newcommand{\GamR}{\Gamma_{\rm R}}

\newcommand{\la}{\langle}
\newcommand{\ra}{\rangle}

\begin{document}

 \title{Full counting statistics of level renormalization in electron
 transport through double quantum dots}

 \author{JunYan Luo}\email{jyluo@zust.edu.cn}
 \affiliation{School of Science, Zhejiang University of Science
 and Technology, Hangzhou 310023, China}
 \author{HuJun Jiao}
 \affiliation{Department of Physics, Shanxi University, Taiyuan, 
 Shanxi 030006, China}
 \author{Yu Shen}
 \affiliation{School of Science, Zhejiang University of Science
 and Technology, Hangzhou 310023, China}
 \author{Gang Cen}
 \affiliation{School of Science, Zhejiang University of Science
 and Technology, Hangzhou 310023, China}
 \author{Xiao--Ling He}
 \affiliation{School of Science, Zhejiang University of Science
 and Technology, Hangzhou 310023, China}
 \author{Changrong Wang}
 \affiliation{School of Science, Zhejiang University of Science
 and Technology, Hangzhou 310023, China}

\date{\today}

 \begin{abstract}
 We examine the full counting statistics of electron transport
 through double quantum dots coupled in series, with particular
 attention being paid to the unique features originating
 from the level renormalization.
 It is clearly illustrated that the energy renormalization
 gives rise to a dynamical charge blockade mechanism, which
 eventually results in a super--Poissonian noise.
 Coupling of the double dots to an external heat bath leads to
 dephasing and relaxation mechanisms, which are demonstrated
 to suppress the noise in a unique way.
 \end{abstract}


\maketitle

 \section{\label{thsec1}Introduction}

 The manifestation of quantum coherence in finite systems is
 the foundations of mesoscopic physics.
 Double quantum dots \cite{Wie031,Bra05315}, due to their inherent quantum
 coherence, are widely accepted as promising candidates for
 building scalable qubits \cite{Los98120,Vri00012306,Fri03121301}
 and quantum states detectors \cite{Jia07155333,Gil06116806} towards
 the realization of quantum computation \cite{Nie00}.
 A great deal of effort has been invested to coherently measure,
 characterize and manipulate the quantum states in a double dot
 structure via coupling to a classic field or external
 surrounding \cite{Ono021313,Hay03226804,Pet052180,Kop06766,%
 Fuj061634,Gus082547,Sas09121926}.

 It is well--known that a quantum system loses coherence
 due to coupling with a noisy environment. The involving
 issues have been the subject of intense research for many
 years \cite{Leg871,Wei08,Yan05187}.
 Yet there is another important consequence of the system--environment
 coupling which renormalizes the internal energy of a quantum system.
 It has been revealed that, in a spin valve structure, the energy
 renormalization provides as an effective exchange magnetic
 field which leads to spin
 precession \cite{Kon03166602,Bra06075328,Sot10245319}.
 For a quantum dot Aharonov--Bohm interferometer, the level
 renormalization gives rise to additional dephasing of the
 quantum state \cite{Mar03195305}, as well as bias dependent
 phase shift and asymmetric interference patterns \cite{Bru96114}.
 Even in solid--state quantum state measurement, the level
 renormalization was shown to play essential roles, and
 influence the measurement effectiveness
 crucially \cite{Luo09385801,Luo104904}.
 It is therefore of vital importance to take this feature
 into account in order to correctly understand and analyze the
 electron transport properties.

 To have a specific example, we will investigate the energy
 renormalization of electron transport through a serial double
 quantum dot system, as schematically shown in \Fig{Fig1}.
 The analysis is based on the full counting statistics (FCS),
 which is capable of characterizing the correlations between
 charge transport events of all orders \cite{Lev964845,Bag03085316}.
 It thus serves as an essential tool superior to the average
 current in distinguishing various transport mechanisms
 involved \cite{Bla001,Naz03}.
 We demonstrate unambiguously that the level renormalization
 gives rise to a dynamical charge blockade mechanism, which
 eventually results in a pronounced super--Poissonian noise.
 Close relation between the super--Poissonian noise and the
 negative differential conductance (NDC) owing to the level
 shift are revealed.
 It is further illustrated that the noise can be strongly
 suppressed due to coupling with an external phonon bath.

 The paper is structured as follows. In \Sec{thsec2} we present
 the model Hamiltonian for the double dot transport system.
 Section\,\ref{thsec3} is devoted to the theory of FCS.
 The bias voltage dependence of the level renormalization and its
 influence on the FCS are discussed in \Sec{thsec4}, which is then
 followed by the summary in \Sec{thsec5}.

 \section{\label{thsec2}Model description}

 \begin{figure}
 \begin{center}
 \includegraphics*[scale=0.8]{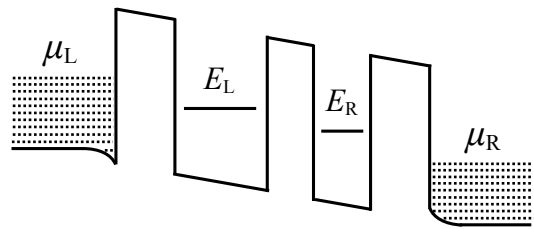}
 \caption{\label{Fig1}Schematic illustration for transport through
 serial double quantum dots.}
 \end{center}
 \end{figure}

 The system under study is schematically shown in \Fig{Fig1},
 where the double dots are coupled in series.
%%%%%
 The entire system is described by the Hamiltonian
 \be\label{Htot}
 H=H_\rmB+H_\rmS+H'.
 \ee
 The first part $H_\rmB=\sum_{\alf=\rmL,\rmR}\sum_{k,\sgm}\vpl_{\alf k}
 c_{\alf k\sgm}^\dag c_{\alf k\sgm}$
 models the noninteracting electrons in the
 electrodes. Here $c_{\alf k\sgm}$
 ($c_{\alf k\sgm}^\dag$) denotes the annihilation (creation)
 operator for electrons in the left ($\alf$= L) or right ($\alf$= R) electrode
 with spin $\sgm=\upa$ or $\dwa$.
 The electron reservoirs are assumed to be in local equilibrium,
 so they are characterized by the Fermi functions
 $f_{\rmL/\rmR}(\omg)$.
 Electron transport is achieved by applying a bias
 voltage $V$, which is dropped symmetrically at the
 left and right tunnel junctions. The bias
 then is modeled by different chemical
 potentials in the left and right electrodes
 $\mu_{\rmL/\rmR}=\pm V/2$.
 Throughout this work, we set $e=\hbar=1$ for electron charge and
 the Planck constant, unless stated otherwise.

 The second part of the Hamiltonian depicts the coupled dots:
 \bea\label{Hs}
 H_\rmS&=&\sum_{\alf=\rmL,\rmR}\left[\sum_\sgm E_\alf n_{\alf\sgm}
 +U_0 n_{\alf\upa}n_{\alf\dwa}\right]
 +U'n_\rmL n_\rmR
 \nl
 &&+\,\Omg\sum_\sgm(d_{\rmL\sgm}^\dag d_{\rmR\sgm}
 +d_{\rmR\sgm}^\dag d_{\rmL\sgm}).
 \eea
 Here, $n_{\alf\sgm}=d_{\alf\sgm}^\dag d_{\alf\sgm}$ and
 $n_\alf=\sum_\sgm n_{\alf\sgm}$ are the occupation number
 operators for dot $\alf$=L or R, with $d_{\alf\sgm}$
 ($d_{\alf\sgm}^\dag$) being the annihilation (creation)
 operator of an electron in the dot $\alf$ with spin $\sgm$.
 Each quantum dot is assumed to have only one spin-degenerate
 energy level $E_{\rmL/\rmR}$ within the bias window.
 One can parametrize the levels by their average energy
 $\bar{E}=(E_\rmL+E_\rmR)/2$ and their difference
 $\epl=E_\rmL-E_\rmR$, such that $E_{\rmL/\rmR}=\bar{E}\pm\frac{1}{2}\epl$.
 $\Omg$ accounts for the interdot tunneling between the coupled
 dots.
 Simultaneous occupation of one electron in each dot
 is associated with the interdot charge energy $U'$.
 Double occupation on the same dot cost the intradot
 charge energy $U_0$, which is assumed to be much larger
 than the bias voltage, such that charge states with
 three or more electrons in the double dots are prohibited.

 The third part
 $H'=\sum_{\alf,\sgm}(f_{\alf\sgm}d_{\alf\sgm}^\dag+d_{\alf\sgm}f_{\alf\sgm}^\dag)$
 describes electron tunneling between the dots and electrodes, with
 $f_{\alf\sgm}\equiv\sum_k t_{\alf k}c_{\alf k\sgm}$.
 The effects of the stochastic electron reservoirs are
 encapsulated in the correlation functions
 $C^{(+)}_{\alf\sgm\sgm'}(t-\tau)\equiv
 \la f_{\alf\sgm}^\dag(t)f_{\alf\sgm'}(\tau)\ra$ and
 $C^{(-)}_{\alf\sgm\sgm'}(t-\tau)\equiv
 \la f_{\alf\sgm}(t) f^\dag_{\alf\sgm'}(\tau)\ra$. Here,
 $\la(\cdots)\ra\equiv{\rm Tr}[(\cdots)\rho_\rmB]$
 represents the thermal average, with $\rho_\rmB$ the
 local thermal equilibrium reservoir state.
%%%
 Due to the serial geometry, electrons in the left (right)
 electrode can only transfer to the left (right) dot.
 The tunnel coupling strength of electrode $\alf$=L or R
 to the corresponding dot is characterized by the intrinsic line width
 $\Gam_\alf(\omg)=2\pi\sum_k|t_{\alf k}|^2\dlt(\vpl_{\alf k}-\omg)$.
 In what follows, we consider only spin conserving tunneling processes
 and assume flat bands in the electrodes, which yields
 energy independent couplings $\Gam_\alf$.

 \section{\label{thsec3}Full counting statistics}

 The dynamics of the reduced system is described by the reduced
 density matrix $\rho(t)$, which is obtained from the density
 matrix of the entire system by integrating out the reservoir
 degrees of freedom.
 Under the second--order Born--Markov approximation, it satisfies
 the quantum master equation \cite{Yan982721}
 \be\label{QME}
 \dot{\rho}(t)=-\rmi {\cal L}\rho(t)-{\cal R}\rho(t),
 \ee
 where the first part represents the internal dynamics
 on the double dots, with ${\cal L}(\cdots)\equiv[H_\rmS,(\cdots)]$.
 The second term accounts for the tunnel coupling
 between double dots and the external electrodes. Its
 detailed structure will be specified soon.
%%%
 The quantum master equation (\ref{QME}) fully captures the
 dynamics of the reduced system; however, it is not adequate
 to describe the output characteristics.

 We unravel the reduced density matrix $\rho(t)$ into components
 $\rho^{(n)}(t)$, in which ``$n$'' denotes the number of
 electrons that have been transferred to the right electrode. The
 resultant particle--number--resolved master equation
 reads \cite{Li05066803,Li05205304,Luo07085325,Luo08345215},
 \bea\label{CME}
 \dot{\rho}^{(n)}&\!\!\!=\!\!\!&-\rmi {\cal L}\rho^{(n)}
 \!\!-\!\!\frac{1}{2}\!\sum_{\sgm}
 \!\Bigg\{\!\sum_{\alf=\rmL,\rmR}\!\left[d_{\alf\sgm}^\dag A_{\alf\sgm}^{(-)}\!\rho^{(n)}
 \!+\!\rho^{(n)}\!A_{\alf\sgm}^{(+)} d_{\alf\sgm}^\dag\right]
 \nonumber \\
 &\!\!\!&-\Big[\,d_{\rmL\sgm}^\dag\,\rho^{(n)}A_{\rmL\sgm}^{(+)}
 +A_{\rmL\sgm}^{(-)}\rho^{(n)}d_{\rmL\sgm}^\dag+\,
 d_{\rmR\sgm}^\dag\,\rho^{(n+1)}A_{\rmR\sgm}^{(+)}
 \nonumber \\
 &\!\!\!&+\,A_{\rmR\sgm}^{(-)}\rho^{(n-1)}d_{\rmR\sgm}^\dag\Big]
 +\,{\rm H.c.}\Bigg\},
 \eea
 where $A^{(\pm)}_\sgm=\sum_{\sgm}A^{(\pm)}_{\alf\sgm}$,
 and $A^{(\pm)}_{\alf\sgm}\equiv\sum_{\sgm'}
 \{C^{(\pm)}_{\alf\sgm\sgm'}(\pm{\cal L})
 +\rmi D^{(\pm)}_{\alf\sgm\sgm'}(\pm{\cal L})\}
 d_{\alf\sgm'}$.
 Here, the involving spectral functions are defined as the Fourier
 transform of the bath correlation functions, i.e.,
 \be
 C^{(\pm)}_{\alf\sgm\sgm'}(\pm{\cal L})=\int_{-\infty}^\infty
 \!\rmd t\,e^{\pm \rmi{\cal L}t}C^{(\pm)}_{\alf\sgm\sgm'}(t).
 \ee
 The dispersion function $D^{(\pm)}_{\alf\sgm\sgm'}(\pm{\cal L})$ then
 is determined via \cite{Yan05187,Xu029196}
 \be
 D^{(\pm)}_{\alf\sgm\sgm'}(\pm{\cal L})=-\frac{{\cal P}}{\pi}
 \!\int_{-\infty}^\infty \!\!\rmd\omg
 \frac{C^{(\pm)}_{\alf\sgm\sgm'}(\pm\omg)}{{\cal L}-\omg},
 \ee
 with ${\cal P}$ denoting the principal part.
%%%%%
 The spectral functions are associated with particle transfer
 processes, and the dispersion functions account for the
 coupling--induced energy renormalization of the dot levels.
%%%%%
 The latter has been neglected in previous
 work \cite{Gur9615932,Gur9715215,Gur986602,Sto961050,Haz01165313},
 in which the Fermi energies of the electrodes are assumed to be far
 away from the electronic states of the dots.
 Later, it will be shown that the level renormalization can
 give rise to intriguing and important features in the output
 characteristics.

 By summing up all the components $\rho^{(n)}(t)$ in \Eq{CME},
 one recovers the unconditional quantum master equation.
 Consequently, the dissipative term in \Eq{QME} is specified
 explicitly.
 The unique advantage of the particle--number--resolved master
 equation is its capability of establishing a close link between
 the reduced dynamics and the output characteristics.
 By utilizing the conditional master equation (\ref{CME}), FCS
 characteristics can be readily determined, which enable us to
 get access to the complete information of transport.

 Let us start with the particle--number--resolved reduced density
 matrix, which is directly related to the the probability distribution
 $P(n,t_0)$ of having $n$ electrons transferred through the system during
 the counting time $t_0$, i.e., $P(n,t_0)={\rm Tr}\rho^{(n)}(t_0)$,
 where the trace is over the reduced system states.
 The associated cumulant generating function ${\cal F}(\chi)$ is
 defined as
 \be
 e^{-{\cal F}(\chi)}=\sum_{n}P(n,t_0)e^{-\rmi n\chi},
 \ee
 where $\chi$ is the so--called counting field.
 All cumulants of the
 current can be obtained from the generating function by
 performing derivatives with respect to the counting field
 \be\label{cumulants}
 \la I^k\ra=-\frac{1}{t_0}(-\rmi\partial\chi)^k{\cal F}(\chi)|_{\chi=0}.
 \ee
 The first four cumulants are related to the average current,
 the (zero-frequency) current shot noise, the skewness, and the
 kurtosis, respectively.

 To derive the cumulant generating function, we shall make use of
 the $\chi$--space counterpart of the number--resolved reduced
 density matrix
 $\varrho(\chi,t)\equiv\sum_n \rho^{(n)}(t)e^{\rmi n\chi}$.
 Its equation of motion, by employing the conditional master
 equation (\ref{CME}), reads formally
 \be
 \dot{\varrho}(\chi)\equiv{\cal L}_\chi\varrho(\chi),
 \ee
 where ${\cal L}_\chi$ is totally determined by the dynamical
 structure of \Eq{CME}.
 The formal solution can be readily obtained as
 $\varrho(\chi,t_0)=e^{{\cal L}_\chi t_0}\varrho(\chi,0)$.
 Straightforwardly, the cumulant generating function reads
 ${\cal F}(\chi,t_0)=-\ln\{{\rm Tr}\varrho(\chi,t_0)\}$.
 Particularly, in the zero-frequency limit, i.e., the
 counting time $t_0$ is much longer than the time of
 tunneling through the system, the cumulant generating
 function is simplified
 to \cite{Bag03085316,Gro06125315,Fli05475,Kie06033312}
 \be
 {\cal F}(\chi,t_0)=-\lambda_{\rm min}(\chi) t_0,
 \ee
 where $\lambda_{\rm min}(\chi)$ is the minimal eigenvalue
 of ${\cal L}_\chi$ that satisfies
 $\lambda_{\rm min}|_{\chi\rightarrow0}\rightarrow0$.

 \section{\label{thsec4}Level renormalization and FCS analysis}

 \begin{figure*}
 \includegraphics*[scale=0.75]{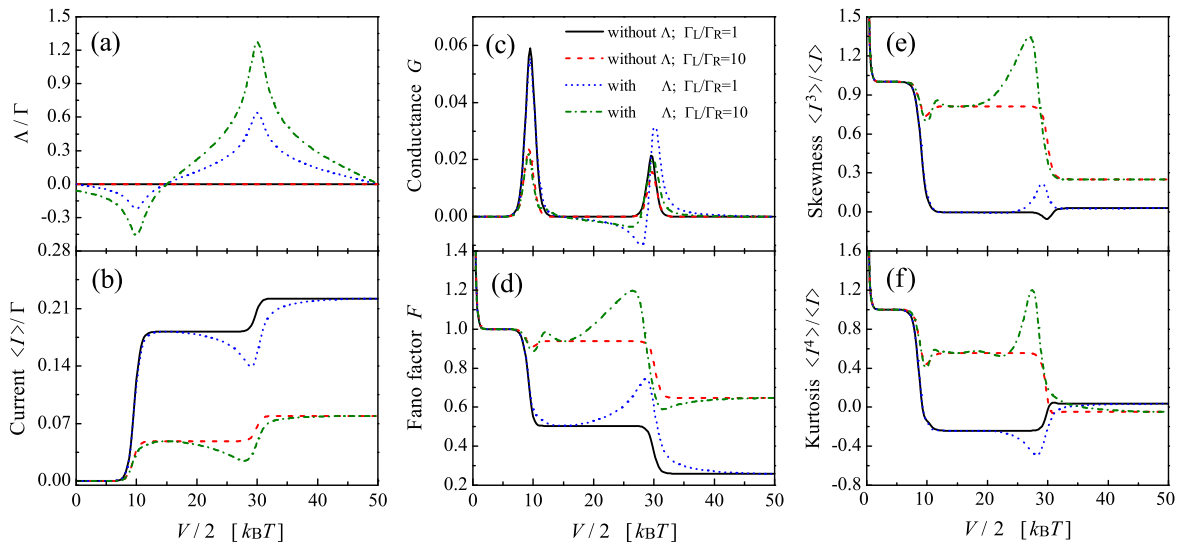}
 \caption{\label{Fig2} (a) level renormalization $\Lambda$, (b) average
 current $\la I\ra$, (c) differential conductance $G=\rmd\langle I\rangle/\rmd V$,
 (d) Fano factor $F=\la I^2\ra/\la I\ra$, (e) normalized skewness, and
 (f) normalized kurtosis
 versus the bias voltage for symmetric ($\GamL=\GamR$) and asymmetric
 ($\GamL=10\GamR$) tunnel couplings.
 The bias is applied symmetrically at the left and right tunnel
 junctions, i.e., raise $\mu_\rmL=V/2$ and lower $\mu_\rmR=-V/2$
 simultaneously.
 Each time when the chemical potential
 of the left electrode aligns with the energy needed for either single ($\bar{E}$)
 or double occupation ($\bar{E}+U'$), the level renormalization reaches
 its local extremum.
 The plotting parameters are: $\epl$=0, $\Gam=\GamL+\GamR=2\Omg$,
 $\bar{E}=10k_\rmB T$, $U'=20k_\rmB T$, and $U_0=100k_\rmB T$.}
 \end{figure*}

 In the strong intradot Coulomb blockade regime, double occupation
 on the same dot is prohibited.
 The involving states are restricted to: $|0\ra$--both dots empty,
 $|\rmL\sgm\ra$--one electron in the left dot,
 $|\rmR\sgm\ra$--one electron in the right dot,
 and $|\rmL\sgm\rmR\sgm'\ra$--one electron
 in each dot, respectively.
 The quantum master equation (\ref{QME}) in this localized state
 representation reads
 \begin{widetext}
 \bsube
 \begin{gather}
 \dot{\rho}_{0}=-2(\Gam_{\rmL}^{+}+\Gam_{\rmR}^{+})\rho_{0}+
 \Gam_{\rmL}^{-}(\rho_{\rmL\upa}+\rho_{\rmL\dwa})
 +\Gam_{\rmR}^{-}(\rho_{\rmR\upa}+\rho_{\rmR\dwa}),
 \\
 \dot{\rho}_{\rmL\sgm}=\rmi \Omg (\rho^{\rmL\sgm}_{\rmR\sgm}-
 \rho^{\rmR\sgm}_{\rmL\sgm})
 -(\GamL^-+2\tilde{\Gam}_{\rmR}^+)\rho_{\rmL\sgm}
 +\GamL^+\rho_0
 +\tilde{\Gam}_{\rmR}^-(\rho_{\rmL\sgm\rmR\sgm}
 +\rho_{\rmL\sgm\rmR\bar{\sgm}}),
 \\
 \dot{\rho}_{\rmR\sgm}=\rmi \Omg (\rho^{\rmR\sgm}_{\rmL\sgm}
 -\rho^{\rmL\sgm}_{\rmR\sgm})
 -(\GamR^-+2\tilde{\Gam}_{\rmL}^+)\rho_{\rmR\sgm}
 +\GamR^+\rho_0
 +\tilde{\Gam}_{\rmL}^-(\rho_{\rmL\sgm\rmR\sgm}
 +\rho_{\rmL\bar{\sgm}\rmR\sgm}),
 \\
 \dot{\rho}_{\rmL\sgm\rmR\sgm'}=
 -(\tilde{\Gam}_{\rmL}^-+\tilde{\Gam}_{\rmR}^-)\rho_{\rmL\sgm\rmR\sgm'}
 +\tilde{\Gam}_{\rmL}^+\rho_{\rmR\sgm'}
 +\tilde{\Gam}_{\rmR}^+\rho_{\rmL\sgm},
 \\
 \dot{\rho}^{\rmL\sgm}_{\rmR\sgm}=\rmi\,\tilde{\epl}\,\rho^{\rmL\sgm}_{\rmR\sgm}
 +\rmi\Omg(\rho_{\rmL\sgm}-\rho_{\rmR\sgm})
 -{\textstyle \frac{1}{2}}(\GamL^-+\GamR^-)\rho^{\rmL\sgm}_{\rmR\sgm}
 -(\tilde{\Gam}_{\rmL}^++\tilde{\Gam}_\rmR^+)\rho^{\rmL\sgm}_{\rmR\sgm},\label{off-diag}
 \end{gather}
 \esube
 \end{widetext}
 with spin $\sgm=\{\upa,\dwa\}$ and $\bar{\sgm}=-\sgm$.
 Here $\rho_s\equiv\la s|\rho|s\ra$ represents the diagonal element
 of the reduced density matrix. The off--diagonal elements
 $\rho^s_{s'}\equiv\la s|\rho|s'\ra$
 describes the so--called quantum ``coherencies''.
 The involving temperature--dependent tunneling rates are defined
 as $\Gam_\alf^{\pm}\equiv\Gam_\alf f_\alf^{(\pm)}(\bar{E})$ and
 $\tilde{\Gam}_\alf^{\pm}\equiv\Gam_\alf f_\alf^{(\pm)}(\bar{E}+U')$,
 where $f^{(\pm)}_\alf(\omg)=\{1+e^{\pm (\omg-\mu_\alf)/k_{\rmB}T}\}^{-1}$
 is related to the Fermi function of the electrode $\alf$= L or R.
 Here we are interested in the regime $\Dlt\ll k_{\rm B}T$
 ($\Dlt=\sqrt{\epl^2+4\Omg^2}$ being the eigenenergy separation),
 where the external coupling strongly modifies the internal
 dynamics, and the off--diagonal elements of the reduced density
 matrix have essential roles to play
 \cite{Gur9615932,Sto961050,Agu04206601}.
 The level separation is thus smeared by the temperature, and only
 excitation energies $\bar{E}$ and $\bar{E}+U'$ enters the Fermi
 functions.

 By a close observation of the off--diagonal element of the
 reduced density matrix [see \Eq{off-diag}], it is found that the
 level detuning $\epl$ is renormalized to
 \be
 \tilde{\epl}=\epl+\Lambda,
 \ee
 with $\Lambda=\Lambda_\rmL-\Lambda_\rmR$ the energy renormalization.
 The level shift $\Lambda_\alf$ arising from tunnel coupling to the
 electrode $\alf$ is given by
 \be\label{renormalization}
 \Lambda_\alf=\phi_\alf(\bar{E})-2\phi_\alf(\bar{E}+U')
 +\phi_\alf(\bar{E}+U_0),
 \ee
 with
 \be
 \phi_\alf(\omg)=\frac{\Gam_\alf}{2\pi}{\rm Re}\left[\Psi
 \left(\frac{1}{2}+\rmi \frac{\omg-\mu_\alf}{2\pi k_\rmB T}\right)\right].
 \ee
 Here $\Psi$ is the digamma function. The involving intradot
 charging energy $U_0$ serves as a natural cut--off for the
 energy renormalization \cite{Wun05205319}.

 The level renormalization is a genuine interaction effect; it
 vanishes for $U_0=U'=0$.
 To clearly elucidate the effect of energy renormalization $\Lambda$,
 hereafter we assume $\epl=0$.
 The numerical result of $\Lambda$ is plotted in \Fig{Fig2}(a) as
 a function of bias voltage.
 Each time when the Fermi energy of the left electrode is resonant
 with the energy needed for single ($\bar{E}$) or double
 occupation ($\bar{E}+U'$), the level shift reaches a local
 extremum.
 The energy renormalization is sensitive to the tunnel--coupling
 asymmetry, i.e., $\Lambda$ increases with rising $\GamL/\GamR$ ratio.

 The level renormalization gives rise to unique features
 in the transport current, as shown in \Fig{Fig2}(b).
 By neglecting the level shift, the current shows a typical
 step--like structure (see the solid and dashed curves). That is,
 each time when a new electronic level enters the bias
 window defined by the chemical potentials of the left
 and right electrodes, a current step occurs.
 The renormalized level detuning leads to suppression of the
 current (see the dotted and dash--dotted curves), particularly
 on the second current plateau, where at most one electron can
 reside on the double dots due to strong interdot and
 intradot charging energies.
 In this double--dot Coulomb blockade (DDCB)
 regime \cite{Luo07085325,Luo08345215}, an analytical
 expression for the transport current can be obtained
 readily by utilizing \Eq{cumulants},
 \be
 \la I\ra = \frac{\GamR\Omg^2}
 {\Omg^2(2+\GamR/2\GamL)+\GamR^2/4+\Lambda^2},
 \ee
 where the involving Fermi functions are approximated
 by either one or zero.
 Apparently, whenever the magnitude of the level detuning
 renormalization grows, the current reduces (the suppression of current
 close to $\bar{E}$ can be hardly resolved due to weak level
 renormalization and finite temperature).
%%%
 The suppression of the current leads to regimes of NDC, as
 shown in \Fig{Fig2}(c). Actually, NDC has also been observed in
 many different contexts \cite{San08035409,Wan02125307}.

 Double occupation on the system (one electron in each dot)
 becomes energetically allowed when the chemical potential of the
 left electrode crosses the excitation level $\bar{E}+U'$.
 The current rises to the third plateau, which corresponds to
 the single--dot Coulomb blockade regime.
 The stationary current is given by
 \be
 \la I\ra =\frac{4(2\GamL+\GamR)\GamL\GamR\Omg^2}
 {\GamL\GamR[(2\GamL+\GamR)^2+4\Lambda^2]+2(2\GamL+\GamR)^2\Omg^2}.
 \ee
 Again the current is suppressed whenever the level renormalization
 grows [see also the dotted and dash--dotted curve in \Fig{Fig2}(b)].

 Let us now consider the second cumulant, which characterizes
 the width of the current distribution and is directly
 related to the shot noise.
 Commonly, it can be expressed in terms of the
 so--called Fano factor $F=\la I^2\ra/\la I\ra$.
 The numerical result against the bias voltage is plotted
 in \Fig{Fig2}(d) for symmetric and asymmetric tunnel couplings.
 At small bias ($V\ll k_{\rm B}T$), the thermal noise dominates,
 and is described by the well--known hyperbolic cotangent
 behavior which leads to a divergence of the Fano
 factor at $V=0$.
%%%
 As bias increases but still well blow $\bar{E}$, electron
 transport is exponentially suppressed. Tunneling events are
 uncorrelated and the noise exhibits Poissonian
 statistics \cite{Bla001}.
 The transport through the system becomes energetically allowed
 when the bias is further increased to the DDCB regime.
 Owing to the level renormalization, the Fano factor exhibits
 clear enhancement at bias close to the excitation energies
 $\bar{E}$ and $\bar{E}+U'$.
 Hence the Fano factor proves to be much sensitive to the internal
 energy than the average current.

 Noticeably, a pronounced super--Poissonian noise is observed
 in the DDCB regime [see the dash--dotted curve in \Fig{Fig2}(d)].
 By evaluating the minimal eigenvalue of ${\cal L}_\chi$, one
 obtains from \Eq{cumulants} the analytical expression of the
 Fano factor
 \be\label{Fano}
 F=1-4\GamL \Omg^2\frac{\GamR(\GamR^2+6\GamL\GamR+8\Omg^2)+
 4\Lambda^2(\GamR-2\GamL)}
 {[2\Omg^2(4\GamL+\GamR)+\GamL(\GamR^2+4\Lambda^2)]^2}.
 \ee
 Unambiguously, super--Poissonian noise is expected when the
 second term is negative. This is satisfied under the conditions
 of $\Lambda\neq0$ and $\GamR<2\GamL$. In this case, the
 tunnel--coupling to the right electrode is
 not very strong, and thus the Coulomb interactions are more
 effective.

 \begin{figure}
 \begin{center}
 \includegraphics*[scale=0.75]{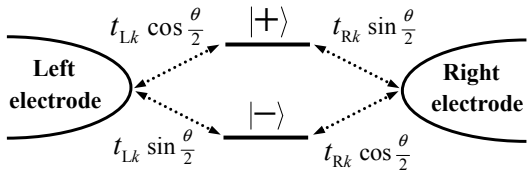}
 \caption{\label{Fig3}Schematic parallel two--level system obtained
 by a unitary transform of the double quantum dots coupled in serial
 as shown in \Fig{Fig1}.
 The tunneling amplitudes are effectively modulated by the level
 renormalization.}
 \end{center}
 \end{figure}

 In the DDCB regime, states with two or more electrons in the
 double dots are prohibited.
 We perform a unitary transformation to diagonalize the reduced
 system Hamiltonian as (spin indices are suppressed here in order
 to simplify the discussion)
 \be
 {\tilde H}_\rmS=\frac{1}{2}\tilde{\Delta}(|+\ra\la+|-|-\ra\la-|),
 \ee
 where $\tilde{\Dlt}=\sqrt{\Lambda^2+4\Omg^2}$, with the level
 renormalization being properly accounted for.
 The serial double dots are thus mapped onto a parallel
 two--level system, as schematically shown in \Fig{Fig3}.
 Here, the energy eigenstates are defined as
 \bsube
 \begin{align}
 |+\ra&\equiv\sin\frac{\theta}{2}|\rmR\ra+\cos\frac{\theta}{2}|\rmL\ra,
 \;\; \textrm{(bonding state)}
 \\
 |-\ra&\equiv\cos\frac{\theta}{2}|\rmR\ra-\sin\frac{\theta}{2}|\rmL\ra,
 \;\; \textrm{(anti--bonding state)}
 \end{align}
 \esube
 where  $\theta$ is introduced via $\sin\theta=2\Omg/\tilde{\Dlt}$ and
 $\cos\theta=\Lambda/\tilde{\Dlt}$.
 As a result, the system--electrode coupling Hamiltonian
 is recast to
 \begin{align}
 {\tilde H}'=&\sum_k\left\{\left(t_{\rmL k}
 \cos\case{\theta}{2} c^\dag_{\rmL k}
 +t_{\rmR k}\sin\case{\theta}{2}
 c^\dag_{\rmR k}\right)|0\ra\la+|\right.
 \nonumber \\
 &\quad\,+\left.\left(t_{\rmR k}\cos\case{\theta}{2} c^\dag_{\rmR k}
 -t_{\rmL k}\sin\case{\theta}{2}
 c^\dag_{\rmL k}\right)|0\ra\la-|\right\}+{\rm H. c.}.
 \end{align}
 Apparently, the effective amplitudes (as shown in \Fig{Fig3}) of
 electron tunneling through the bonding and anti--bonding states
 are affected notably by the level renormalization.

 For $\Lambda=0$ ($\theta=\pi/2$) and symmetric tunneling amplitudes
 ($t_{\rmL k}=t_{\rmR k}$), the noise of electron transport through
 the bonding and anti--bonding states is suppressed due to the Pauli
 exclusion principle. It leads thus to a sub--Poissonian statistics,
 as shown by the solid curve in \Fig{Fig2}(d).
 Finite tunnel--coupling asymmetry enhances the degree of correlation
 in transport. Yet, the noise can not exceed the Poisson value by
 increase the tunnel--coupling asymmetry alone [cf. \Eq{Fano}].
 Correlation of transport can be crucially increased by the
 level renormalization.
 As the level shift $\Lambda$ grows, electron transport
 through the bonding and anti--bonding states strongly modulate
 each other, and a dynamical channel blockade mechanism is
 developed \cite{Luo08345215,Kie07206602,Luo10083720,Cot04206801,San07146805}.
 It gives rise to the bunching of tunneling events, and
 eventually results in the super--Poissonian noise, as
 shown by the dash--dotted curve in \Fig{Fig2}(d).

 The occurrence of the dynamic charge blockade leads generally
 to the suppression of the current, which finally gives rise
 to the NDC.
 However, the NDC does not necessary imply the super--Poissonian
 noise, as shown by the dotted curves in \Fig{Fig2}(c) and (d).
 This seems to be at variance with that in Ref. \cite{Thi05045341},
 where electron transport through a multi--level quantum dot is
 investigated. There the NDC is found to be always accompanied with
 super--Poissonian noise.
 NDC was also observed in Ref. \cite{Wan02125307}, where the
 occurrence of NDC was associated with reduction of charge
 accumulation due to  bias dependent tunneling rates.
 The NDC in the present case is a pure energy
 renormalization effect, as will be explained below.

 For the present double dot system, the rates of tunneling through
 the bonding and anti--bonding states (as shown in \Fig{Fig3}) depend
 not only on the tunneling amplitude ($t_{\alf k}$), but also on the
 level renormalization $\Lambda$.
 In the case $t_{\rmL k}=t_{\rmR k}$, enhancement of the level
 renormalization results in a decrease of $\sin\frac{\theta}{2}$.
 The rates of electron tunneling from left electrode to the
 anti--bonding state and that from bonding state to the right
 electrode are both suppressed.
 As a result, the current is reduced and NDC occurs [see the dotted
 curve in \Fig{Fig2}(c)].
 However, the noise does not exceed the Poissonian value due to
 suppressed transport through both channels.
 Now, consider an increase of $t_{\rmL k}$, which enhances the
 rate of electron tunneling from the left electrode to the
 double dots.
 In the presence of a strong energy renormalization, the rate of
 electron tunneling from the bonding state to the right electrode
 remains very low.
 In the limit where the Coulomb interactions prevent a double
 occupancy of system, there is competition between the two transport
 channels. Consequently, the slow flowing of electrons through
 the bonding state modulates that through the anti--bonding state,
 which gives rise to a bunching of tunneling events and
 eventually leads to the super-Poissonian noise and NDC, as
 displayed by the dash--dotted curves in \Fig{Fig2}(c) and (d).

 We are now in a position to examine the third and fourth
 cumulants, which characterize respectively the asymmetry
 and sharpness of the current distribution. The normalized
 skewness is displayed in \Fig{Fig2}(e).
 In the DDCB regime, it exhibits notable enhancement due to
 energy renormalization. In particular, a pronounced
 super--Poissonian behavior is observed for strongly asymmetric
 tunnel couplings, as displayed by the dash--dotted curve in
 \Fig{Fig2}(e).
 The kurtosis, as shown in \Fig{Fig2}(f), can be either enhanced
 or reduced by the energy renormalization, depending sensitively
 on the tunnel coupling asymmetry.
 The peak width of kurtosis, in comparison with that of Fano factor
 and skewness, is reduced obviously; it thus reflects more precisely
 where the resonance and the extremum of energy renormalization are
 achieved.

 \begin{figure}
 \begin{center}
 \includegraphics*[scale=1.2]{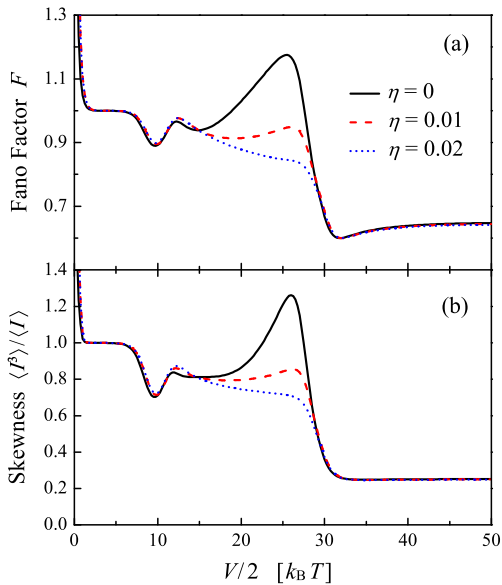}
 \caption{\label{Fig4}(A) Fano factor and (b) normalized skewness
 versus bias voltage for different dissipative couplings $\eta$.
 The tunnel couplings to the left and right electrodes are
 asymmetric ($\GamL=10\GamR$) and the energy cutoff for the
 phonon bath spectral density is $\omg_{\rm c}=5\Gam$. The other
 parameters are the same as those in \Fig{Fig2}.}
 \end{center}
 \end{figure}

 Finally, let us turn to the influence of external phonon
 bath which leads to relaxation and dephasing in the
 system \cite{Bra05315,Wan05121303}. The
 corresponding rates are given respectively
 by \cite{Agu04206601,Kie07206602,Luo10083720}
 \bsube
 \be
 \Gam_{\pm}= -\frac{\pi}{2}\frac{\Omg}{\Dlt}J_{\rm ph}(\Dlt)
 \left[\frac{\epl}{\Dlt}\coth\left(\frac{\Dlt}{2k_\rmB T}\right)\pm1\right]
 +\eta\pi\frac{k_\rmB T\Omg\epl}{\Dlt^2},
 \ee
 and
 \be
 \gamma_{\rm ph}=2\pi\frac{\Omg^2}{\Dlt^2}J_{\rm ph}(\Dlt)
 \coth\left(\frac{\Dlt}{2k_\rmB T}\right)+\eta\pi\frac{k_\rmB T\epl^2}{\Dlt^2},
 \ee
 \esube
 where $J_{\rm ph}(\omg)=\eta \omg e^{-\omg/\omg_{\rmc}}$ is the
 Ohmic spectral density of the heat bath, with the dimensionless
 parameter $\eta$ reflecting the strength of dissipation and
 $\omg_{\rm c}$ the Ohmic high energy cutoff.

 The calculated Fano factor and normalized skewness versus
 bias voltage are plotted in \Fig{Fig4} for different
 dissipative couplings $\eta$.
 Strong suppression of the cumulants is observed in the
 DDCB regime, where double occupation on the  system is
 prohibited.
 In the absence of electron--phonon coupling, electrons are
 transferred coherently. The phonon bath coupling, on
 one hand, leads to dephasing
 of the quantum state, and electrons tend to be
 tunneled sequentially, which then causes the reduction of
 noise in the DDCB regime \cite{Kie07206602}.
 On the other hand, the involving phonon emission and
 absorption processes give rises to a relaxation
 mechanism, which further suppresses the noise. The
 noise suppression is particularly notable
 in the bias close to $\bar{E}+U'$, where the energy
 renormalization is prominent, as shown by the dotted
 curves in \Fig{Fig4}(a) and (b) for a strong
 dissipative coupling $\eta=0.02$.
 These features demonstrate that both Fano factor and
 skewness are sensitive tools to the phonon bath
 induced dephasing  and relaxation.

 \section{\label{thsec5}Conclusion}
 In summary, we have investigated the full counting statistics of
 electron transport through double quantum dots coupled in series,
 and paid particular attention to the unique features arising from
 the level renormalization.
%%%%
 It is found that a dynamical channel blockade mechanism is
 developed purely due to the energy renormalization, which eventually
 leads to a pronounced super--Poissonian noise.
 Our results demonstrate unambiguously the importance of the level
 renormalization when investigating the transport properties of a
 double dot structure.
 Negative differential conductance due to level detuning renormalization
 is observed, and its relation with the super--Poissonian noise is
 revealed.
 Coupling of the double dots to an external phonon bath leads to
 dephasing and relaxation mechanisms, which are shown to suppress
 noise notably.
 Furthermore, double dot systems, as recently shown in Refs. 
 \cite{Yin05L183,Ema09161309}, are good candidates for
 entanglement generation.
 In this context, the present full counting statistics have the
 potential to facilitate the identification and characterization
 of entanglement originating from different sources.

 \begin{acknowledgments}
 Support from the National Natural Science Foundation of China
 (Grant Nos. 10904128 and 11004124),
 and the Natural Science Foundation of Zhejiang Province
 (grant nos Y6100171 and Y6110467) are gratefully acknowledged.
 \end{acknowledgments}

%\bibliography{E:/bibliography/bibrefs}
%\bibliographystyle{unsrt}
%\bibliography{E:/bibliography/bibrefs}

\end{document}